\documentstyle[preprint,aps]{revtex}
\tightenlines

\begin{document}
\newcommand{\beq}{\begin{equation}}
\newcommand{\eeq}{\end{equation}}
\newcommand{\beqa}{\begin{eqnarray}}
\newcommand{\eeqa}{\end{eqnarray}}
\newcommand{\sr}{\sqrt}
\newcommand{\fr}{\frac}
\newcommand{\mn}{\mu \nu}
\newcommand{\G}{\Gamma}

\draft
\preprint{ INJE-TP-00-05}
\title{U(1) Gauge Field of the Kaluza-Klein Theory  \\
       in the Presence of Branes}
\author{ Gungwon Kang\footnote{E-mail address:
kang@physics.inje.ac.kr} and Y.S. Myung\footnote{E-mail address:
ysmyung@physics.inje.ac.kr} }
\address{
Department of Physics, Inje University,
Kimhae 621-749, Korea}
\maketitle
\begin{abstract}

We investigate the zero mode dimensional reduction of the
Kaluza-Klein unifications in the presence of a single brane in the
infinite extra dimension. We treat the brane as fixed, not a
dynamical object, and do not require the orbifold symmetry. It seems
that, contrary to the standard Kaluza-Klein models, the 4D effective
action is no longer invariant under the U(1) gauge transformations
due to the explicit breaking of isometries in the extra dimension by
the brane. Surprisingly, however, the linearized perturbation analysis
around the RS vacuum shows that the Kaluza-Klein gauge field does
possess the U(1) gauge symmetry at the linear level. In addition,
the graviscalar also behaves differently from the 4D point of view.
Some physical implications of our results are also discussed.

\end{abstract}
\bigskip

\newpage

\section{Introduction}

In the standard five-dimensional Kaluza-Klein (KK) approach where
the full spacetime manifold is factorized as $M_4 \otimes S^1$,
the five-dimensional gravitation theory which has the reparametrization
invariance on $S^1$ can be interpreted as a gauge theory of the
Virasoro group from the four-dimensional point of view (see
\cite{CZ} and references therein). After the geometric spontaneous
symmetry breaking of the Virasoro invariance, the excitations of the
5D gravitational fields are split into 4D massless gravitational
fields, massless gauge fields, massless scalar field, and an
infinite tower of massive spin-2 fields~\cite{CZ,SSRSS}.
In particular, the U(1) gauge symmetry of the vector fields in
the 4D effective action is originated from the translational isometries
in the extra dimension.

Recently, there have been lots of interest in the phenomenon of
localization of gravity proposed by Randall and Sundrum (RS)~\cite{RS2,RS1}
(for previous relevant work see references therein). RS~\cite{RS1} assumed
a single positive tension 3-brane and a negative bulk cosmological
constant in the five-dimensional spacetime. There have been developed
a large number of brane world models afterwards~\cite{CEHS,OMs}.
The introduction of branes usually gives rise to the ``warping" of the
extra dimensions, resulting in non-factorizable spacetime manifolds.
More importantly, the presence of branes breaks the translational
isometries in the extra dimensions. Therefore, it would be very interesting
to see how the conventional Kaluza-Klein picture changes in the brane world
scenarios.

Dobado and Maroto~\cite{DM} recently have incorporated the effect of the
presence of the brane in the Kaluza-Klein (KK) reduction by introducing
Goldstone bosons (GB) fields. The GB fields parameterize the excitations
of the brane~\cite{GB} and so the GB correspond to the spontaneous
symmetry breaking of the translational isometries of the compactified
extra dimensions by the brane. It has been shown that in the dimensional
reduction of the KK unifications a sort of Higgs mechanism related to the
GB~\cite{bHiggs} gives mass to the KK graviphotons. (see also the
appendix of Ref.~\cite{DKS} although it has the opposite sign for the
mass term.)

On the other hand, there are other approaches to confine standard model
particles on the brane by allowing the fields to live in the bulk
spacetime. For example, bulk gauge bosons have been considered in
Ref.~\cite{BulkFs,DFKK}. It is important to derive the zero mode
effective action because its zero modes (massless modes) correspond
to the standard model particles localized on the brane.

In this paper, contrary to the approach mentioned above where the brane
is treated as a dynamical object, we have treated the brane as fixed,
and investigate the zero mode dimensional reduction of the Kaluza-Klein
unifications. The brane world model considered in this paper is the RS one
with a single 3-brane in the infinite fifth dimension~\cite{RS1}.
As expected, the breaking of isometries in the extra dimension by the
brane makes the 4D effective action not being invariant under U(1) gauge
transformations. Interestingly, however, the analysis of the linearized
equation around the RS background shows that the KK gauge field does
possess the U(1) gauge symmetry at the linear level.

In section 2, we carry out the dimensional reduction of the KK unifications
in the presence of a single brane with some ansatz for the zero mode
excitations. In section 3, the linearized perturbations of the 4D
effective action obtained are analyzed around the RS vacuum solution.
Some physical implications of our results are discussed in section 4.

\section{Kaluza-Klein reduction}

The RS model with a single brane can be described by the following
action in 5D spacetimes~\cite{RS1,CEHS,RSmodels}
\beq
I = \int d^4x \int^{\infty}_{-\infty} dz
    \fr{\sr{-\hat{g}}}{16\pi G_5} (\hat{R} -2\Lambda )
    - \int d^4x \sr{-\hat{g}_B} \sigma .
\label{5DI}
\eeq
Here $G_5$ is the 5D Newton's constant, $\Lambda$ the bulk cosmological
constant of 5D spacetime, $\hat{g}_B$ the determinant of the induced
metric describing the brane, and $\sigma$ the tension of the brane.
We assume that the value of $\sigma$ is fine-tunned such that
$\Lambda =-6k^2 (< 0)$ with $k=4\pi G_5 \sigma /3$. Let us introduce
the domain-wall metric in the following form
\beqa
ds^2 &=& \hat{g}_{MN} dx^Mdx^N = H^{-2}(z) g_{MN} dx^Mdx^N
\nonumber \\
&=& H^{-2}(z) \Big[ \gamma_{\mu\nu}dx^{\mu}dx^{\nu}
    +\phi^2 (dz -\kappa A_{\mu}dx^{\mu})^2 \Big] .
\label{metric}
\eeqa
Here $H= k|z|+1$, $\phi^2 = g_{55}$, and $\kappa A_{\mu}=-g_{5\mu}
/g_{55}$. The standard Kaluza-Klein decomposition of the metric is
given by
\beq
(g_{MN}) = \left(\matrix{ 
\gamma_{\mu\nu}+\kappa^2 \phi^2 A_{\mu}A_{\nu} & -\kappa \phi^2 A_{\mu}\cr 
-\kappa \phi^2 A_{\nu} & \phi^2 \cr}
\right) ,
\qquad
(g^{MN}) = \left(\matrix{ 
\gamma^{\mu\nu} & \kappa A^{\mu} \cr 
\kappa A^{\nu} & \phi^{-2}(1+\kappa^2\phi^2 A\cdot A) \cr} 
\right)
\label{KKm}
\eeq
with $A^{\mu} = \gamma^{\mu\nu} A_{\nu}$ and $A\cdot A = A_{\mu}A^{\mu}$.
Here $\kappa$ is the gauge coupling constant.

Under the specific class of coordinate transformations such as
\beq
x^{\mu} \rightarrow \tilde{x}^{\mu} =  \tilde{x}^{\mu} (x), \qquad
z \rightarrow \tilde{z} = z + f(x),
\label{CT}
\eeq
we obtain
\beq
\tilde{\gamma}_{\mu\nu} = \fr{\partial x^{\alpha}}{\partial
\tilde{x}^{\mu}} \fr{\partial x^{\beta}}{\partial \tilde{x}^{\nu}}
\gamma_{\alpha\beta}, \qquad \tilde{A}_{\mu} = \fr{\partial
x^{\alpha}}{\partial \tilde{x}^{\mu}} A_{\alpha} + \kappa^{-1}
\fr{\partial f}{\partial \tilde{x}^{\mu}},
\qquad \tilde{\phi} (\tilde{x}, \tilde{z}) = \phi (x,z)
\eeq
according to $\tilde{g}_{MN} = \fr{\partial x^P}{\partial
\tilde{x}^M}\fr{\partial x^Q}{\partial \tilde{x}^N} g_{PQ}$.
We see that $\gamma_{\mu\nu}$ transforms like a four-dimensional
metric tensor, and $\phi$ a scalar field under diffeomorphisms in
Eq.~(\ref{CT}). However, we point out that the 5D diffeomorphisms are
split into the 4D diffeomorphisms plus the gauge transformations for
the field $A_{\mu}$.

In this paper, we are mainly interested in the zero mode effective
action . In general, it is a non-trivial problem to
determine what the ``zero mode" is if the full spacetime is not
factorizable. As an ansatz for the zero mode, we assume that
$\gamma_{\mu\nu}$, $A_{\mu}$, and $\phi$ are functions of
$x$-coordinates only, i.e., no $z$-coordinate dependence.
If one requires the $Z_2$ (e.g., $R/Z_2$) orbifold symmetry in the 
brane world model, there will be no vector zero mode fluctuations. 
It follows because in the presence of $Z_2$ orbifold symmetry 
in Eq.~(\ref{metric}) the vector gauge field $A_{\mu}$ should satisfy 
$A_{\mu}(x, -z) = -A_{\mu}(x, z)$ and so $A_{\mu}(x)=0$. In what 
follows, we consider general cases without having the orbifold 
symmetry in the theory.

The above assumption comes from the crucial observation that 
the graviton zero mode $h_{\mn}$ in $\gamma_{\mn}=\eta_{\mn} 
+\kappa h_{\mn}$ depends only on ``$x$" even if one starts from 
$h_{\mn}(x,z)=H^{3/2} \psi(z)\hat h_{\mn}(x)$ in the RS 
approach~\cite{RS1} where the transverse fluctuations are fixed 
(e.g., $h_{5\mu} = h_{55} =0$). For the zero mode solution with 
$m^2=0$, we have $\psi^0(z)= c_h H^{-3/2}$, thus we find 
$h^0_{\mn}(x,z)= c_h \hat h_{\mn}(x)$ with a constant $c_h$. 
Other examples are the form of the zero
modes for the bulk spin-0 and spin-1 fields on the RS background~\cite{BG}.
For the spin-0 field $\Phi(x,z)=H^{3/2} \chi(z)\hat \phi(x)$, we
have $\chi= c_\Phi H^{-3/2}$ for the zero mode and hence its
localized zero mode is given by $\Phi^0(x,z)= c_\Phi \hat
\phi(x)$. In the case of the spin-1 field $V_\mu(x,z)= H^{3/2} 
\sigma(z) v_\mu(x)$, one finds $\sigma= c_V H^{-3/2}$ for the zero 
mode and hence its zero mode is given by $V^0_\mu(x,z)= c_V v_\mu(x)$. 
From the observations mentioned above, we may propose that the zero 
modes are constants with respect to ``$z$".
Furthermore we stress that for $h_{\mu5} \not=0, h_{55} \not=0$,
it may be not a correct way to obtain the zero modes from the linearized 
equations. This is because their forms are too complicated to analyze 
the zero modes~\cite{RSmodels}. Even if we choose the gauge-fixing, 
it is hard to obtain the consistent zero mode solutions. Hence the 
integration of Eq.~(\ref{5DI}) over $z$ could be a good starting point 
to obtain the zero modes.

Note first that
\beqa
\mbox{} && \sr{-\hat{g}} = H^{-5}\phi \sr{-\gamma} , \\
&& \sr{-\hat{g}_{B}} =H^{-4}(z=0) \sr{-\gamma} \sr{|\delta^{\mu}_{\nu}
+\kappa^2 \phi^2 A^{\mu}A_{\nu}|} .
\eeqa
Using $\hat{g}_{MN} = H^{-2}g_{MN}$, one has
\beq
\hat{R} = H^2 \Big[ R(g) +8\fr{\nabla_P\nabla^PH}{H}
-20\fr{\nabla_PH\nabla^PH}{H^2} \Big].
\eeq
Since
\beq
\nabla_P\nabla^PH = H''(\phi^{-2}+\kappa^2 A\cdot A) +\kappa H'
    (\phi^{-1}A^{\mu}\partial_{\mu}\phi +\fr{1}{2}
    A^{\mu}\gamma^{\alpha\beta}\partial_{\mu}\gamma_{\alpha\beta}
    +\partial_{\mu}A^{\mu}) ,
\eeq
we have
\beqa
8\fr{\nabla_P\nabla^PH}{H}-20\fr{\nabla_PH\nabla^PH}{H^2}
&=& \Big( 8\fr{H''}{H} -20(\fr{H'}{H})^2 \Big)
  (\phi^{-2}+\kappa^2 A\cdot A)  \nonumber  \\
&& +8\kappa \fr{H'}{H}
  \big( \phi^{-1}A^{\mu}\partial_{\mu}\phi +\fr{1}{2}
  A^{\mu}\gamma^{\alpha\beta}\partial_{\mu}\gamma_{\alpha\beta}
  +\partial_{\mu}A^{\mu} \big) ,
\eeqa
where the prime ($'$) denotes the differentiation with respect to
$z$.

Then, the five-dimensional action Eq.~(\ref{5DI}) is given by
\beqa
I &=& \fr{1}{16\pi G_5} \int d^4x \sr{-\gamma} \phi
  \Big[ R(g) \int dz H^{-3} +(\phi^{-2} +\kappa^2 A\cdot A)
  \int dz H^{-3} \Big( 8\fr{H''}{H} -20\fr{H'^2}{H^2} \Big)
  \nonumber \\
&& +8\kappa \big( \phi^{-1}A^{\mu} \partial_{\mu}\phi +\fr{1}{2}
  A^{\mu}\gamma^{\alpha\beta} \partial_{\mu}\gamma_{\alpha\beta}
  +\partial_{\mu}A^{\mu} \big) \int dz \fr{H'}{H^4}
  -2\Lambda \int dz H^{-5} \Big]  \nonumber   \\
&& -\int d^4x \sr{-\gamma} \sr{|\delta^{\mu}_{\nu}
  +\kappa^2 \phi^2 A^{\mu}A_{\nu}|} \sigma .
\eeqa
It should be pointed out that the 5D Ricci scalar curvature
constructed from $g_{MN}$, $R(g)$, is independent of $z$-coordinate
since the metric elements $g_{MN}$ are functions of $x^{\mu}$ only.
Using $H' =k\theta (z)$, $H''=2k\delta (z)$, $\int^{\infty}_{-\infty}
dz H^{-3} = 1/k$, and $\int^{\infty}_{-\infty} dz H^{-5} = 1/2k$
for the RS model with a single brane, one gets the 4D effective action
\beqa
\label{BKKI}
I_{KK} &=& \fr{1}{16\pi G_5} \fr{1}{k} \int d^4x \sr{-\gamma}
  \Big[ \phi R(g) -\phi \Lambda +6k^2 (\phi^{-1}
  +\kappa^2 \phi A\cdot A)   \nonumber  \\
&& \qquad\qquad\qquad\qquad\qquad\qquad\qquad -16\pi G_5 k\sigma
  \sr{|\delta^{\mu}_{\nu} +\kappa^2 \phi^2 A^{\mu}A_{\nu}|} \Big] \\
&=& \fr{1}{16\pi G_4} \int d^4x \sr{-\gamma}\Big[ \phi R(g)
  +6k^2 \Big( \phi^{-1} +\phi -2\sr{|\delta^{\mu}_{\nu} +\kappa^2
  \phi^2 A^{\mu}A_{\nu}|} +\kappa^2\phi A\cdot A \Big) \Big] .
\label{KKI0}
\eeqa
Here the 4D Newton's constant is defined as $G_4=G_5 k$.
Notice from Eq.~(\ref{BKKI}) that it reproduces the ordinary KK
reduction in the presence of the cosmological constant as the brane
at $z=0$ disappears (i.e., $\sigma$, $k \rightarrow 0$).
The 5D scalar curvature $R(g)$ is related to the 4D Ricci scalar
curvature constructed from $\gamma_{\mu\nu}$, $R(\gamma )$, as
\beq
\int d^4x \sr{-\gamma}\phi R(g) = \int d^4x \sr{-\gamma}
  \phi \Big[ R(\gamma ) -\fr{\kappa^2}{4}\phi^2 F^2 \Big]
  +\oint [\cdots ]  .
\label{5R4R}
\eeq
The last term is the surface term. The field strength is defined
as $F_{\mu\nu}= \partial_{\mu}A_{\nu} -\partial_{\nu}A_{\mu}$ and
$F^2 =F_{\mu\nu} F^{\mu\nu}$. Using Eq.~(\ref{5R4R}), one finally
obtains
\beqa
I_{KK} &=& \fr{1}{16\pi G_4} \int d^4x \sr{-\gamma}\Big[ \phi
  R(\gamma ) -\fr{\kappa^2}{4} \phi^{3} F^2  \nonumber  \\
&& \qquad\qquad\qquad\qquad +6k^2 \Big( \phi^{-1} +\phi
  -2\sr{|\delta^{\mu}_{\nu} +\kappa^2 \phi^2 A^{\mu}A_{\nu}|}
  +\kappa^2\phi A\cdot A \Big) \Big] ,
\label{KKI1}
\eeqa
where we have ommitted the surface terms.

We observe that the zero mode gravitational degrees of freedom
in the 5D spacetime are split into the 4D gravitational fields
$\gamma_{\mu\nu}$, a vector field $A_{\mu}$, and a graviscalar
 field $\phi$ as usual. However, the properties of the
vector field and the scalar field are very different from those
in the conventional KK reduction. The first two terms in this
effective action are the same form as in the ordinary dimensional
reduction of the Kaluza-Klein unifications, and they have the U(1)
gauge symmetry. The difference from the conventional KK reduction
is only the last term which is proportional to the brane tension
squared. If one started from the KK metric decomposition with
$A_{\mu}=0$ and $\phi =1$ in Eq.~(\ref{KKm}), this ``potential"
term would disappear and one obtains the ordinary Einstein gravity
on the brane with zero effective cosmological constant as well known.
As can be easily seen in Eq.~(\ref{BKKI}), this happens because of
the fine tunning between the brane tension $\sigma$ and the 5D bulk
cosmological constant $\Lambda$.

The appearance of the non-linear
term (e.g., $\sr{|\delta^{\mu}_{\nu} +\kappa^2 \phi^2
A^{\mu}A_{\nu}|}$) as well as the squared term in $A_{\mu}$ shows
not only that the 4D effective action no longer has the gauge
symmetry, but also that KK photons are not massless. This arises from
the presence of the brane ($k \neq 0$) in the five-dimensional
spacetime. Because Eq.~(\ref{KKI1}) contains a non-linear term which
generates a lot of terms, a truncated form of the effective action
may be useful to understand the dynamics of the fields easily. If one
expands the full action up to the order of $\kappa^2$, one has
\beq
I^T_{KK} \simeq \fr{1}{16\pi G_4} \int d^4x \sr{-\gamma}\Big[ \phi
  R(\gamma ) -\fr{\kappa^2}{4} \phi^{3} F^2
  +6k^2 \Big( \phi^{-1} +\phi -2
  +\kappa^2 \phi (1 -\phi ) A\cdot A \Big) \Big] .
\label{KKI1appr}
\eeq
Here we used $\sr{|\delta^{\mu}_{\nu} +\kappa^2 \phi^2 A^{\mu}A_{\nu}|}
\simeq 1+\fr{1}{2} \kappa^2 \phi^2 A\cdot A$.

In order to explicitly see how the dynamical aspect of the $\phi$
field comes out, let us conformally transform the metric as
\beq
\gamma_{\mu\nu} \rightarrow \bar{\gamma}_{\mu\nu} =\phi \gamma_{\mu\nu}.
\eeq
Then, the zero-mode effective action Eq.~(\ref{KKI1}) is written by
\beqa
I_{KK} &=& \fr{1}{16\pi G_4} \int d^4x \sr{-\bar{\gamma}}\Big[
  R(\bar{\gamma}) -\fr{\kappa^2}{4} \phi^{3} F^2
  -\fr{3}{2} \phi^{-2} \bar{\nabla}^{\mu}\phi \bar{\nabla}_{\mu}\phi
   \nonumber  \\
&& \qquad\qquad  +6k^2 \phi^{-2}\Big( \phi^{-1} +\phi
  -2\sr{|\delta^{\mu}_{\nu} +\kappa^2 \phi^3 A^{\mu}A_{\nu}|}
  +\kappa^2\phi^2 A\cdot A  \Big) \Big]
\label{KKI2}
\eeqa
up to the surface terms. Here $F^2= \bar{\gamma}^{\mu\nu}
\bar{\gamma}^{\alpha\beta} F_{\mu\alpha}F_{\nu\beta}$, $A^{\mu}
= \bar{\gamma}^{\mu\alpha} A_{\alpha}$, and $A\cdot A =
\bar{\gamma}^{\mu\nu} A_{\mu}A_{\nu}$. The truncated effective
action in this Einstein frame is also given by
\beqa
I^T_{KK} & \simeq & \fr{1}{16\pi G_4} \int d^4x \sr{-\bar{\gamma}}
  \Big\{ R(\bar{\gamma}) -\fr{\kappa^2}{4} \phi^{3} F^2
  -\fr{3}{2} \phi^{-2} \bar{\nabla}^{\mu}\phi \bar{\nabla}_{\mu}\phi
  \nonumber  \\
&& \qquad\qquad  +6k^2 \Big[ \phi^{-2}(\phi^{-1} +\phi -2)
  +\kappa^2 (1-\phi ) A\cdot A  \Big] \Big\}  .
\label{KKI2appr}
\eeqa
As shall be shown below, the form of these actions above also suggests
that the metric tensor which is responsible for the 4D Einstein
gravitation is not the $\gamma_{\mu\nu}$, but indeed the
$\bar{\gamma}_{\mu\nu}$.

Now the field equations derived from the effective action (\ref{KKI2appr})
are given by
\beqa
\mbox{} && R_{\mu\nu} = \fr{\kappa^2}{2}\phi^3 \Big( F_{\mu\alpha}
  {F_{\nu}}^{\alpha} -\fr{1}{4}\bar{\gamma}_{\mu\nu}F^2 \Big)
  -6k^2\kappa^2 (1-\phi )A_{\mu}A_{\nu}   \nonumber  \\
&& \qquad\qquad\qquad  +\fr{3}{2}\phi^{-2}
  \Big[ \bar{\nabla}_{\mu}\phi \bar{\nabla}_{\nu}\phi
  -2k^2 \bar{\gamma}_{\mu\nu} (\phi^{-1}+\phi -2) \Big] ,
\label{EOMg}   \\
&& \bar{\nabla}^{\mu}F_{\mu\nu} +12k^2\kappa^2\phi^{-3}(1-\phi )A_{\nu}
  = -3\phi^{-1} \bar{\nabla}^{\mu}\phi F_{\mu\nu} ,
\label{EOMa}  \\
&& \bar{\nabla}^{\mu}\bar{\nabla}_{\mu} \phi
  -\phi^{-1}\bar{\nabla}^{\mu}\phi \bar{\nabla}_{\mu}\phi
  +2k^2 \Big[ \phi^{-1}(4-\phi -3\phi^{-1}) -\kappa^2\phi^2 A\cdot A
  \Big] = \fr{\kappa^2}{4} F^2 .
\label{EOMp}
\eeqa
Here we used the truncated effective action for simplicity.
The essential properties of solutions do not change.

It is well known in the RS model that if the flat metric on the brane
is replaced by any Ricci-flat 4D metric then the 5D Einstein equations
with a negative cosmological constant are still satisfied 
\cite{EJM,CHR}. Now note that $A_{\mu} =0$ and $\phi =1$ satisfies
Eqs.~(\ref{EOMa}) and (\ref{EOMp}). In this case, Eq.~(\ref{EOMg})
becomes $R_{\mu\nu}=0$. Thus any 4D Ricci-flat metric
$\bar{\gamma}_{\mu\nu}$ is a solution. Therefore, although based on 
the assumption for the zero-mode which is $z$-coordinate independent, 
our results reproduce this well known property in the RS model. 
Such somewhat general agreement may indicates the validity of our 
ansatz for the zero mode. One can see that $\bar{\gamma}_{\mu\nu} 
=\eta_{\mu\nu}$ corresponds to the RS solution~\cite{RS1}. 
Since the metric for the 4D Schwartzschild black hole is Ricci-flat,
the Schwartzschild black hole can be embedded in the brane world as 
well known. That is, assuming the spherically symmetric background
$\hat g_{MN}=H^{-2}(z)(\bar \gamma^S_{\mn},\phi^2)$ with 
$\bar{\gamma}^S_{\mu\nu} =
{\rm diag} [ 1-2M/r, (1-2M/r)^{-1}, r^2, r^2\sin^2\theta ]$ and $\phi
=1$, we obtain the black string solution in 5D Anti de Sitter($AdS_5$)
spacetime~\cite{CHR}. Since this black string solution is unstable near 
the AdS horizon, but stable far from it, it is likely to end up with 
a ''black cigar" solution as conjectured in Ref. \cite{CHR}. 

Secondly, it would be of interest to ask whether or not the
Reissner-Nordstr\"{o}m charged black hole can be embedded in the
brane world. It is straightfoward to see that
the Reissner-Nordstr\"{o}m black hole solution,
\beq
\bar{\gamma}^{RN}_{\mu\nu} = {\rm diag} [ 1-2M/r+Q'^2/r^2,
(1-2M/r+Q'^2/r^2)^{-1}, r^2, r^2\sin^2\theta ]
\eeq
with $Q'=\kappa Q/2$, $F_{tr}=Q/r^2$, and $\phi =1$, satisfies
Eqs.~(\ref{EOMg}) and (\ref{EOMa}). In this case we have $F_{tr}= 
-\partial_r A_0$ with $A_0= Q/r$. However it does not satisfy
Eq.~(\ref{EOMp}) (i.e., $-2k^2A\cdot A = F^2/4$). This is so 
because of the presence of both brane and $\phi$. Thus, the 4D
Reissner-Nordstr\"{o}m black hole with $\phi =1$ cannot be embedded
in the brane world. In the absence of the brane (i.e., $k=0$), there
exist charged black hole solutions with non-trivial $\phi$ field.
However, such charged black hole is very different from the
Reissner-Nordstr\"{o}m black hole~\cite{HHCM}. It does not seem that
the presence of the brane changes this feature of the Kaluza-Klein
theory much. On the other hand, however, if the scalar field were
frozen somehow from the beginning in Eq.~(\ref{KKm}) (e.g., $\phi =1$),
there would be no equation like Eq.~(\ref{EOMp}). Consequently, by
observing Eqs.~(\ref{EOMg}) and (\ref{EOMa}), one can easily find that
the Reissner-Nordstr\"{o}m black hole is a solution.

\section{Linearized perturbation}

In this section we consider the linearized perturbations around the
RS vacuum solution ($\bar{\gamma}_{\mu\nu}=\eta_{\mu\nu}, A_{\mu}=0,
\phi =1$) for the dimensionally reduced effective action in
Eq.~(\ref{KKI2}). Actually, at the linear level, the truncated effective
action in Eq.~(\ref{KKI2appr}) is equivalent to the non-truncated one
in Eq.~(\ref{KKI2}). Let us introduce the perturbations around the
RS solution
\beq
\gamma_{\mu\nu} = \eta_{\mu\nu} + \kappa h_{\mu\nu},
\qquad A_{\mu}=0 +a_{\mu}, \qquad  \phi = 1+\kappa \varphi .
\eeq
Consequently,
\beq
\bar{\gamma}_{\mu\nu} = \eta_{\mu\nu} +\kappa \bar{h}_{\mu\nu},
\qquad\qquad  \bar{h}_{\mu\nu} = h_{\mu\nu} +\varphi \eta_{\mu\nu}.
\eeq
Then the bilinear action of Eq.~(\ref{KKI2}) or (\ref{KKI2appr}) which
governs the perturbative dynamics is given by
\beqa
I^{0}_{KK} &=& \fr{\kappa^2}{16\pi G_4} \int d^4x \Big\{- \fr{1}{4}\Big[
  \partial^{\mu}\bar{h}^{\alpha\beta}\partial_{\mu}\bar{h}_{\alpha\beta}
  -\partial^{\mu}\bar{h}\partial_{\mu}\bar{h} +2\partial^{\mu}\bar{h}_{\mu\nu}
  \partial^{\nu}\bar{h} -2\partial^{\mu}\bar{h}_{\mu\alpha}\partial^{\nu}
  {\bar{h}_{\nu}}^{\alpha} \Big]   \nonumber  \\
&& \qquad\qquad -\fr{1}{4}(\partial_{\mu}a_{\nu} -\partial_{\nu}a_{\mu})
  (\partial^{\mu}a^{\nu} -\partial^{\nu}a^{\mu})
  -\fr{3}{2}\partial_{\mu}\varphi \partial^{\mu}\varphi
  +6k^2 \varphi^2 \Big\},
\label{KKILP}
\eeqa
where $\bar{h}=\eta^{\mu\nu}\bar{h}_{\mu\nu}=h+4\varphi$. Surprisingly,
it turns out that the bilinear effective action is invariant under the
U(1) gauge transformation. The U(1) gauge symmetry breaking term in
Eq.~(\ref{KKI2appr}) appears as higher order term than the squared order:
i.e., $6k^2\kappa^2 (1-\phi ) A\cdot A \simeq -6k^2\kappa^3 \varphi
a^{\mu}a_{\mu}$.

In order to understand what physical states there are, let us analyze
the field equations as below. From the action Eq.~(\ref{KKILP}) we have
the equations of motion
\beqa
\mbox{} && \Box \bar{h}_{\mu\nu} +\partial_{\mu}\partial_{\nu}
  \bar{h} -\Big( \partial_{\mu}\partial^{\alpha}\bar{h}_{\alpha\nu}
  +\partial_{\nu}\partial^{\alpha}\bar{h}_{\alpha\mu} \Big)
  -\eta_{\mu\nu} \Big( \Box \bar{h} -\partial^{\alpha}\partial^{\beta}
  \bar{h}_{\alpha\beta} \Big) = 0 ,
\label{spin2}  \\
&& \Box a_{\mu} -\partial_{\mu} (\partial_{\nu}a^{\nu}) =0,
\label{spin1}  \\
&& \Box \varphi +4k^2 \varphi = 0.
\label{spin0}
\eeqa

By taking the trace of Eq.~(\ref{spin2}), we have
\beq
\Box \bar{h} -\partial^{\alpha}\partial^{\beta} \bar{h}_{\alpha\beta}
  = 0  .
\label{Trace}
\eeq
Hence Eq.~(\ref{spin2}) becomes
\beq
\Box \bar{h}_{\mu\nu} +\partial_{\mu}\partial_{\nu}
  \bar{h} -\Big( \partial_{\mu}\partial^{\alpha}\bar{h}_{\alpha\nu}
  +\partial_{\nu}\partial^{\alpha}\bar{h}_{\alpha\mu} \Big)
  = 0 .
\label{spin2b}
\eeq

So far we have not chosen any gauge for $\bar{h}_{\mu\nu}$. Now let us
choose the transverse (or harmonic) gauge in the five-dimensional
spacetime. Since
\beq
g_{MN} = \eta_{MN} +\kappa \epsilon_{MN}, \qquad\qquad
\Big( \epsilon_{MN} \Big) = \left(\matrix{ 
h_{\mu\nu} & -a_{\mu} \cr
-a_{\nu} & 2\varphi \cr}
\right)  ,
\eeq
the five-dimensional harmonic gauge $\partial^M \epsilon_{MN}
=\fr{1}{2} \partial_N \epsilon$ is equivalent to
\beq
\partial^{\mu}\bar{h}_{\mu\nu} =\fr{1}{2}\partial_{\nu} \bar{h},
\qquad\qquad  \partial_{\mu}a^{\mu} =0.
\eeq
That is, the 5D harmonic gauge is split into the harmonic gauge for
the 4D gravitational field and the Lorenz gauge for the 4D KK gauge
field. Using these gauge conditions above, Eq.~(\ref{spin2b}) and
Eq.~(\ref{spin1}) become
\beq
\Box \bar{h}_{\mu\nu} = 0 , \qquad\qquad\qquad
\Box a_{\mu} = 0 ,
\label{spin2c}
\eeq
respectively. Therefore, it proves that $\bar{h}_{\mu\nu}$ and
$a_{\mu}$ indeed represent the massless spin-2 gravitons and the
massless spin-1 graviphotons on the brane, respectively.

On the other hand, the spin-0 scalar field fluctuation $\varphi$ in
Eq.~(\ref{spin0}) appears to be massive. However, it has a tachyonic
mass $m^2_{\varphi} = -4k^2$ proportional to the brane tension
squared. It seems to indicate that the scalar fluctuation of the 5D
gravitational degrees of freedom corresponds to the unstable mode
in the RS background.

\section{Discussion}

We have investigated the KK zero mode effective action in the presence
of a single brane in the extra dimension. Although the four-dimensional
gravitational modes behaves as usual, the vector and scalar modes behave
quite differently. In the 4D effective action, it seems that the vector
field $A_{\mu}$ does not possess the U(1) gauge symmetry and that KK
photons are not massless any more. The scalar field $\phi$ is also no
longer massless and couples to the vector field.

However, this is not all of the story.
The linearized perturbation analysis around the RS background spacetime 
shows that the 5D massless gravitational degrees of freedom are split 
into spin-2, spin-1, and spin-0 modes as the standard KK model up to 
$\kappa^2$-order. We have observed that the massive propagation of the
vector mode in the 4D effective action is not revealed in the linearized
perturbation. In order to observe the effect of the brane ($k \neq 0$),
thus, one needs to study one-loops correction rather than the
linearized one. For example, we expect the relevant vertex correction
($k^2\kappa^3\varphi a^{\mu}a_{\mu}$) from the last term of
Eq.~(\ref{KKI2appr}).

We also have observed that the spin-0 mode
propagation has a tachyonic mass, indicating some instability of the
RS background spacetime through the ``radion" or ``modulus" field.
Presumably, it suggests that some stabilization mechanism for
the 55-metric component is necessary in order to have the stable
RS background spacetime with a single brane as in the case of the
two branes. On the other hand, if one requires the $R/Z_2$ orbifold 
symmetry in the brane world model, there will be no vector zero mode 
propagations as mentioned above. Thus, the $R/Z_2$ orbifold symmetry 
with the ``modulus" field stabilization establishes the usual 
localization of gravity on the brane in the RS model~\cite{RS1}.

In deriving the U(1) Maxwell term  from the 5D RS brane model, we use
the conventional Kaluza-Klein approach. Apparently, we find a
non-linear term as well as $A\cdot A$. This arises from a sort of
brane-Higgs effect: Here the isometry of extra dimension was broken
spontaneously by the presence of the brane. Hence we expect that
the gauge field becomes massive. However, we have found that the
massive propagation of the KK gauge field does not reveal at the linear
level. Fortunately, instead we find the massless vector propagation.
What will happen if we take a $z$-coordinate dependent or other form of
ansatz for the ``zero modes"? We still expect there should be 
non-linear or mass terms in the reduced effective action due to the
broken isometry in the extra dimension. However, in order to answer 
whether or not the gauge field becomes massless when linearized, 
some further work in detail is needed. There were  other attempts 
to achieve the 4D U(1) symmetry from the 5D U(1) bulk gauge 
fields~\cite{BulkFs,Gai}.

On the other hand, the RS solution can be extended to accommodate the
Schwarzschild black hole solution on the brane as a zero mode solution.
This is possible because the RS solution is Ricci-flat. Hence the 
Ricci-flat Schwarzschild solution can be embedded into the brane world 
by introducing the spherically symmetric spacetime. Now it is very 
important to check whether or not the RS brane world allows to have 
the Reissner-Nordstr\"{o}m black hole on the brane. As is shown above, 
the Reissner-Nordstr\"{o}m black hole cannot be embedded in the brane 
world, because this case of $A_0 \not =0$ cannot be a solution to the 
effective action of Eq.~(\ref{KKI1}) including the non-linear term and 
$A \cdot A$. To obtain this black hole on the brane, it seems to be 
necessary to introduce some U(1) bulk gauge field in the 
five-dimensional spacetime whose dynamics is localized on the 
brane~\cite{CRSS}.

We have observed that the naive propagations of the scalar field gives 
rise to the tachyonic mass proportional to the tension of the brane. 
It may induce the instability of the RS vacuum. In the linearized
gravity, by using the residual gauge freedom, one can also impose 
the traceless gauge in a source-free region in addition to the 
harmonic gauge. It follows mainly because ``$\Box {\rm Trace}$" vanishes 
in a {\it source-free} region. What will happen if such traceless gauge 
is imposed in our linearized analysis? The five-dimensional trace is 
$\epsilon = \eta^{MN} \epsilon_{MN} = h+2\varphi =\bar{h} -2\varphi$. 
Here $\bar{h} = h+4\varphi$ is used. By combining the trace of 
Eq.~(\ref{spin2c}) and Eq.~(\ref{spin0}), we have
\beq
\Box_{(5)} \epsilon = \Box \bar{h} -2\Box \varphi
=8k^2 \varphi ,
\label{5DTrace}
\eeq
where $\Box_{(5)}=\Box +\partial^2_5$.
Thus, imposing the five-dimensional traceless gauge 
(i.e., $\epsilon =0$) directly results in $\varphi =0$, that is, 
no graviscalar fluctuation. Since $h=\epsilon -2\varphi$ and 
$\bar{h}=\epsilon +2\varphi$, it also means the four-dimensional 
traceless gauge (i.e., $h=\bar{h}=0$). 
In other words, we notice  that the existence of the tachyonic
graviscalar fluctuation is mutually inconsistent with imposing 
the traceless gauge condition. Therefore, the resolution of 
the instability of the RS background spacetime due to the graviscalar 
transforms to whether or not one can impose the traceless gauge. 
In the linearized gravity, the trace of metric fluctuations can be 
set to be zero by using remaining gauge freedom provided that there 
is no matter source in the region in consideration~\cite{MTW}. 
In our analysis, however, since the graviscalar field $\varphi$ plays 
like a matter source in the trace equation above, it does not seem to 
be plausible imposing such gauge condition in the first place.
Presumably such traceless gauge condition can be imposed on $\bar{h}$
($\Box \bar{h} =0$), but not on $h$ ($\Box h = 16k^2 \varphi$). 

As mentioned above, another possible caveat of our result is that
the tachyonic graviscalar fluctuation is merely an artifact of the 
ansatz for the zero mode we used in this paper. For instance, 
instead of $H(z)= k|z|+1$ in Eq.~(\ref{metric}), let us assume 
$H(x,z)= k|z| \phi (x)+1$ for the zero mode. This ansatz is analogous 
to the form used in Ref.~\cite{GW} for the case of RS model with two 
branes. Then the 4D effective action is given by
\beqa
I_{KK} &=& \fr{1}{16\pi G_4} \int d^4x \sr{-\gamma}\Big[ 
  R(\gamma ) -\fr{\kappa^2}{4} \phi^{2} F^2 + 2\phi^{-2}
  \gamma^{\mu\nu} \partial_{\mu}\phi \partial_{\nu}\phi \nonumber  \\
&& \qquad\qquad\qquad\qquad +6k^2 \Big( 2 
  -2\sr{|\delta^{\mu}_{\nu} +\kappa^2 \phi^2 A^{\mu}A_{\nu}|}
  +\kappa^2\phi^2 A\cdot A \Big) \Big] \nonumber  \\ 
& \simeq & \fr{1}{16\pi G_4} \int d^4x \sr{-\gamma}\Big[ 
  R(\gamma ) -\fr{\kappa^2}{4} \phi^{2} F^2 + 2\phi^{-2}
  \gamma^{\mu\nu} \partial_{\mu}\phi \partial_{\nu}\phi \Big] .
\eeqa
Here we see that the graviscalar fluctuation as well as that of the 
gravivector becomes massless in the linearized perturbations. 
Therefore, in order to clarify the issues discussed above,
further investigation is required on what the correct form of the ansatz
is for the zero mode in the brane world scenarios.

Finally, it will be interesting to extend our study to various types
of brane world models~\cite{OMs} as well as to the RS model with two
positive and negative tension branes~\cite{RS2} and see how the U(1)
gauge field behaves in the 4D effective action. It will be also worth
investigating how graviphotons and graviscalar particles interact with
the 5D bulk standard model particles~\cite{BulkFs,DFKK,Demir} in the
presence of branes.

\section*{Acknowledgments}

The authors thank Hyungwon Lee for helpful discussions and the referee
for crucial comments. GK thanks Seungjoon Hyun for usefull 
conversations. This work was supported by the Brain Korea 21 Programme, 
Ministry of Education, Project No. D-0025.

\end{document}